# Reconfigurable neural spiking in bias field-free spin Hall nano oscillator

Sourabh Manna[1], Rohit Medwal[2], Rajdeep Singh Rawat[1,#]

[1]Natural Science and Science Education, National Institute of Education, Nanyang Technological University, 637616, Singapore.

[2]Department of Physics, Indian Institute of Technology Kanpur, Uttar Pradesh, 208016, India.

\# Correspondence rajdeep.rawat@nie.edu.sg.

**Keywords**: Spin Hall Nano Oscillator, Magnetic anisotropy, Bias field-free auto-oscillation, Spintronic spiking neuron, Micromagnetic simulation.

# ABSTRACT

In this study, we theoretically investigate neuron-like spiking dynamics in an elliptic ferromagnet/heavy metal bilayer-based spin Hall nano oscillator (SHNO) in bias field-free condition, much suitable for practical realization of brain inspired computing schemes. We demonstrate regular periodic spiking with tunable frequency as well as the leaky-integrate-and-fire (LIF) behavior in a single SHNO by manipulating the pulse features of input current. The frequency of regular periodic spiking is tunable in a range of 0.5 GHz to 0.96 GHz (460 MHz bandwidth) through adjusting the magnitude of constant input dc current density. We further demonstrate the reconfigurability of spiking dynamics in response to a time varying input accomplished by continuously increasing the input current density as a linear function of time. Macrospin theory and micromagnetic simulation provide insights into the origin of bias field-free auto-oscillation and the spiking phenomena in our SHNO. In addition, we discuss how the shape anisotropy of the elliptic ferromagnet influence the bias field-free auto oscillation characteristics, including threshold current, frequency and transition from in-plane to out-of-plane precession. The SHNO operates below $10^{12}$ A/m$^2$ input current density and exhibits a large auto-oscillation amplitude, ensuring high output power. We show that the threshold current density can be reduced by decreasing the ellipticity of the ferromagnet layer as well as enhancing the perpendicular magnetic anisotropy. These findings highlight the potential of our bias field-free SHNO in designing power-efficient spiking neuron-based neuromorphic hardware.

## I. INTRODUCTION

The pursuit of more efficient brain-inspired computing systems drives us to explore beyond-conventional computing paradigms, including neuromorphic and reservoir computing[1-3]. Neuromorphic hardware employs artificial neurons to emulate the spiking dynamics observed in biological neural networks. Consequently, the spiking neural network (SNN) composed of multiple neuron-like software units, seamlessly integrates with the energy-efficient architectures of neuromorphic hardware due to its sparse event-driven processing. Several neuromorphic computing approaches abstract neurons as nonlinear oscillators[4-6]. Spintronic oscillators are highly suitable for realization of scalable SNN-based neuromorphic hardware because of their nanoscale size and inherent nonlinearity. Spin torque nano oscillators (STNO) and spin Hall nano oscillators (SHNO) have already emerged as efficient candidates for performing various classification and combinatorial optimization tasks[7-13]. In particular, SHNO devices, relying on the spin Hall effect[14] to generate spin-orbit torque (SOT), are growing interest because of their easy fabrication process, miniature footprint, lower Joule heating and robust nonlinear magnetization dynamics[15-22]. In addition, recently demonstrated bias field-free operation of SHNO has expanded its applicability for designing low power AI hardware[23-26]. SHNO devices can also be tuned to exhibit voltage spikes akin to biological neurons[27,28]. Markovic et al. have demonstrated state-of-art neuron-like spiking dynamics in an easy-plane ferromagnet/heavy metal (FM/HM) based nanoconstriction SHNO that can potentially enable unsupervised learning[29]. In such SHNOs, application of short nanosecond current pulses trigger spiking behavior in magnetization dynamics. However, a regular periodic spiking induced by a constant dc input current, such as Huxley-Hodgkin spikes[30] has not been observed yet in FM/HM bilayer SHNO.

In majority of SNN, the leaky integrate-and-fire (LIF) feature of biological neuron is employed due to its simplicity, enabling easier hardware implementation and large-scale integration. Although the LIF behavior has been demonstrated through domain wall motion[31], magnetization auto-oscillation in synthetic antiferromagnet heterostructure[32] and skyrmion dynamics[33], it remains unexplored in bias field-free SHNO devices. In this work, we demonstrate both the regular periodic spiking and LIF behavior in a single FM/HM bilayer SHNO, consists of an elliptical FM interfaced with a rectangular HM layer in bias field-free condition. We comprehensively investigate the underlying mechanism responsible for the bias field-free auto-oscillation of magnetization and subsequent spiking phenomena. The simple elliptic geometry enables precise analytical calculation of demagnetization coefficients[34]. Consequently, we explore the role of shape-dependent demagnetization field in presence of perpendicular magnetic anisotropy (PMA) in our SHNO, which is essential to optimize the device geometry for achieving higher output power, improved tunability and enhanced quality of the spikes.

The paper is structured as follows: we begin with the detailed micromagnetic simulation methodology of our SHNO system. Exploring bias-free auto-oscillation characteristics, we reveal the in-plane (IP) to out-of-plane (OOP) transition in magnetization precession. A macrospin model then elucidates the origin of bias field-free auto-oscillation in our system and how it is influenced by the shape anisotropy of the FM layer. Moving on, we demonstrate the SOT-induced spiking behavior in the OOP precession mode. We explore how the demagnetization field determines the spiking rate and quality (sharpness of the spikes) in our system. Finally, we demonstrate the leaky-integrate-and-fire (LIF) behavior along with the reconfigurable spiking in response to variable input current, highlighting the potential of bias field-free SHNOs in neuromorphic hardware design.

## II.     MICROMAGNETIC SIMULATION OF BIAS FIELD-FREE SHNO

We designed an SHNO device consisting of a 1.4 nm thin elliptic FM (CoFeB) layer interfaced with a rectangular 5 nm thick layer of HM ($\beta$-W), as shown in figure 1(a). We have chosen $\beta$-W as the HM layer because of its high spin Hall angle[19,35,36]. The lateral dimension of the $\beta$-W layer is large enough as compared to the major and minor axes of the elliptic CoFeB layer. Therefore, the Oersted field generated upon passing a charge current through the $\beta$-W layer, is uniform and unidirectional over the CoFeB layer. The major axis and minor axis of the elliptic CoFeB layer are denoted by $a$ and $b$ respectively as shown in the inset of figure 1(a). We also define the axis ratio as $r = a/b$. Note that all the results presented in this letter are obtained for $r = 3$ unless mentioned otherwise. A Cartesian coordinate system is defined with the x-axis parallel to the major axis and y-axis parallel to the minor axis of the ellipse, whereas the z-axis is normal to the plane of the ellipse as shown in figure 1(a).

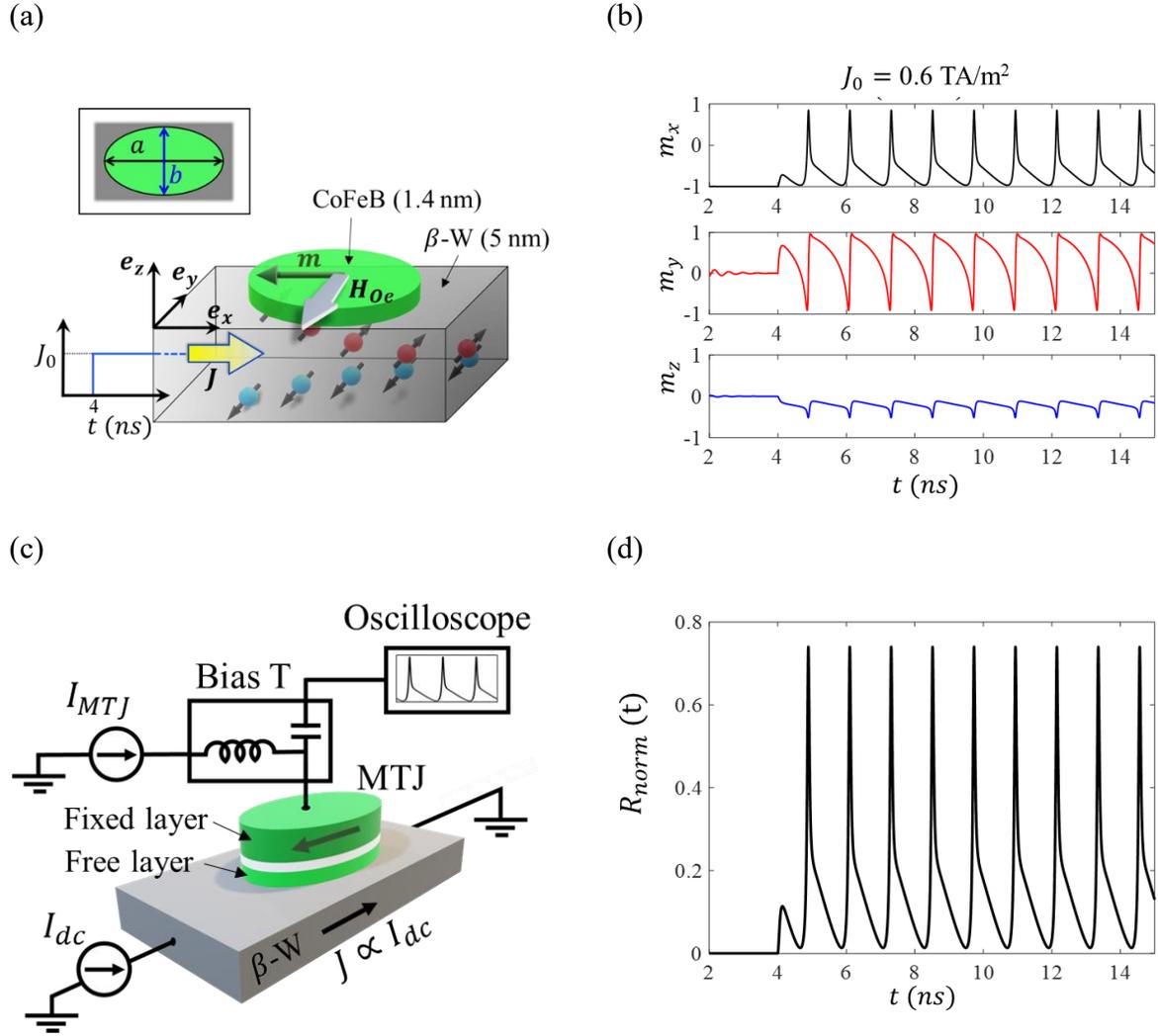

**Figure 1**. (a) Schematic illustration of the SHNO device. $e_x$, $e_y$ and $e_z$ denote the unit vectors along x, y, and z-axes respectively. (Inset) top view of the elliptic FM. (b) Time evolution of the components of normalized magnetization $m$ for input current density, $J_0 = 0.6$ TA/m². (c) Schematic of the electrical circuit for detection of auto-oscillation and spiking dynamics using the TMR of MTJ. The free layer of the MTJ represents the FM layer of (a). (d) Time evolution of the normalized TMR corresponding to the variation of $m$ in (b), calculated using 80% TMR ratio.

The material parameters chosen from experiments in reference[19] are as follows: the CoFeB layer has a saturation magnetization $M_s = 740$ kA/m, exchange constant $A_{ex} = 19$ pJ/m, damping constant $\alpha = 0.023$, PMA field $H_{PMA} = 0.57$ T and gyromagnetic ratio $\gamma = 1.879 \times 10^{11}$ Hz/T. Note that, $H_{PMA}$ is not sufficient to overcome the out-of-plane demagnetization field ($\sim \mu_0 M_s = 0.93$ T). Therefore, the equilibrium magnetization resides in the xy-plane in absence of

SOT. However, the substantial $H_{PMA}$ helps in exciting the magnetization auto-oscillation by reducing the effective out-of-plane demagnetization field. The spin Hall angle of $\beta$-W has been chosen as $\theta_{SH} = -0.41$. A charge current density $J$ along x-axis passing through the $\beta$-W layer, generates a transverse spin current polarized along y-direction. Injection of this spin current into the ferromagnetic CoFeB layer exerts a damping-like SOT (DLT) on the magnetization that counters the damping torque. Note that, the SOT includes a field-like torque (FLT) as well in addition to the damping-like torque in general. However, the FLT in CoFeB/$\beta$-W system is negligible as compared to the DLT[37]. Therefore, we consider only the DLT in the Slonczewski form[38]. The input charge current induces an Oersted field given as $\boldsymbol{H_{Oe}} = -\frac{\mu_0 |J| t_{HM}}{2} \boldsymbol{\hat{y}}$, where $t_{HM}$ denotes the thickness of the HM layer. We emphasize that no external biasing magnetic field has been considered. The magnetization dynamics of the CoFeB layer can then be expressed in terms of the reduced magnetization $\boldsymbol{m} = \boldsymbol{M}/M_s$, following the LLGS equation as[38,39]:

$$\boldsymbol{\dot{m}} = -\gamma \boldsymbol{m} \times \boldsymbol{H_{eff}} + \alpha \boldsymbol{m} \times \boldsymbol{\dot{m}} + \frac{\gamma |J| \hbar \theta_{SH}}{2 e t_{FM} \mu_0 M_s} \boldsymbol{m} \times (\boldsymbol{p} \times \boldsymbol{m}) \qquad (1)$$

In equation (1), $\boldsymbol{H_{eff}} = (\boldsymbol{H_{Oe}} + \boldsymbol{H_{ex}} + \boldsymbol{H_{dem}} + \boldsymbol{H_{PMA}})$ where, $\boldsymbol{H_{ex}}$, $\boldsymbol{H_{dem}}$ and $\boldsymbol{H_{PMA}}$ denote the exchange field, demagnetization field and the PMA field respectively. The spin polarization direction is given by $\boldsymbol{p} = \boldsymbol{e_y}$, where $\boldsymbol{e_y}$ is the unit vector along y-direction. The constants $e, \hbar, t_{FM}$ and $\mu_0$ represent the magnitude of electronic charge, reduced Planck's constant, thickness of the FM layer and permeability of free space respectively.

We have quantitatively obtained the dynamic behavior of magnetization by numerically solving equation (1), using the open-source GPU accelerated software Mumax³[40]. In the simulation, the dc charge current is switched on after 4 ns (i.e., $|J| = 0$ for $t \leq 4\ ns$ and $|J| = J_0$ for $t > 4\ ns$). This delay allows the magnetization to completely relax and reach a stable state before the SOT starts acting on it (see figure 1a). Figure 1(b) shows the time-evolution of the components of $\boldsymbol{m}$ for an input current density, $J_0 = 0.6$ TA/m² in case of $r = 3$. As seen from figure 1(b), the magnetization undergoes sustained auto-oscillation which generate periodic spiking in $m_x, m_y$ and $m_z$ components. However, the spiking amplitude is much more prominent in $m_x$ and $m_y$ as compared to the $m_z$-component, denoting the magnetization precession about a highly out-of-plane axis. In addition, we observe that the amplitude of auto-oscillation is very large in the $xy$- plane that ensures high output power from this bias field-free SHNO. The spikes in $m_x$ can be easily recorded as voltage signal using the TMR of a magnetic tunnel junction (MTJ) and implementing the circuit shown in figure 1(c). The TMR pillar can be designed such that the FM layer adjacent to the HM is the free layer and the top FM layer is the fixed layer with magnetization oriented along the major axis of ellipse. The shape anisotropy of the elliptical fixed layer favors this orientation of magnetization. Hence, no extra antiferromagnetic layer would be required for pinning the magnetization in the fixed layer. The TMR voltage is given as $V = I_{MTJ} \times R_{MTJ}$, where $R_{MTJ}$ is the resistance across the MTJ. We define the normalized $R_{MTJ}$ as[6]:

$$R_{norm} = \frac{r_{TMR}(1 + m_x)}{2} \tag{2}$$

Here, $r_{TMR}$ denotes the TMR ratio given by $r_{TMR} = (R_{MTJ}^{\uparrow\downarrow} - R_{MTJ}^{\uparrow\uparrow})/R_{MTJ}^{\uparrow\uparrow}$. Considering a typical MTJ with 80% TMR, we calculate the time evolution of the normalized TMR, $R_{norm}$ (figure 1d) that shows the periodic spiking behavior as observed in $m_x(t)$.

### III. CHARACTERISTICS OF THE AUTO-OSCILLATION

Looking into the auto-oscillation characteristics, we find that the spiking behavior is not present right from the onset of auto-oscillation. In fact, at lower current density, the magnetization undergoes auto-oscillation about an in-plane (IP) precession axis. Figure 2(a) shows the trajectory of $\boldsymbol{m}$ in such a typical IP precession mode for $J_0 = 0.45$ TA/m$^2$. The corresponding FFT of $m_x(t)$ in figure 2(b) shows the presence of several harmonics, although the 1st harmonic is relatively much stronger than the others. However, $m_x(t)$ does not show spiking behavior in this IP precession (see the inset of figure 2b). Once the $J_0$ exceeds a certain critical value $J_c$, the auto-oscillation occurs about an out-of-plane (OOP) precession axis. Figure 2(c) depicts such an OOP precession trajectory for $J_0 = 0.5$ TA/m$^2$. Note that the 1st harmonic in the corresponding FFT of $m_x(t)$, is shifted towards the higher frequency (see figure 2d). This clearly shows the signature of OOP precession mode[41]. In addition, we observe spiking behavior in $m_x(t)$ in this OOP precession mode, as can be seen from the inset of figure 2(d). The frequency of the 1st harmonic varies nonlinearly as a function of the input current density $J_0$, as depicted in figure 2(e). Initially, this frequency decreases to a certain minimum value at $J_0 = 0.465$ TA/m$^2$, and then starts increasing again, indicating the transition from IP to OOP precession. Hence, the critical current for our system is denoted as $J_c = 0.465$ TA/m$^2$. Moreover, we find that the periodicity of the spiking is governed by the 1$^{st}$ harmonic of auto-oscillation frequency in the OOP precession mode. It is noteworthy that we did not include any thermal effect to understand the intrinsic magnetization dynamics of our SHNO. However, from practical perspective, the effect of thermal fluctuation is important to realize the potential for room temperature operation of such bias field-free SHNO. Hence, we carried out the same simulations at 300 K, considering a random thermal field [40,42]. We observe a similar nonlinear behavior of the 1$^{st}$ harmonic as a function of $J_0$, although the weaker higher harmonics are suppressed by the thermal noise (see supplementary figure S1).

(a)
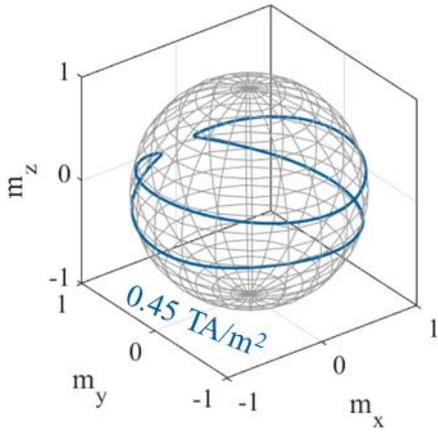

(b)
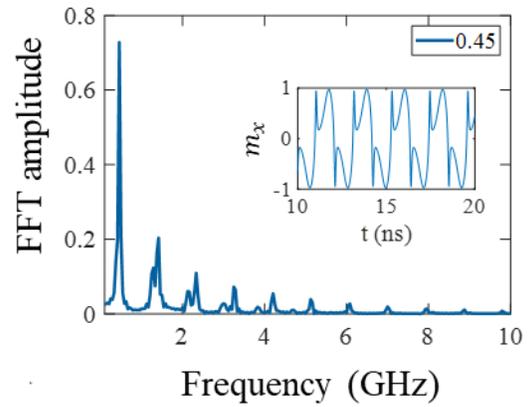

(c)
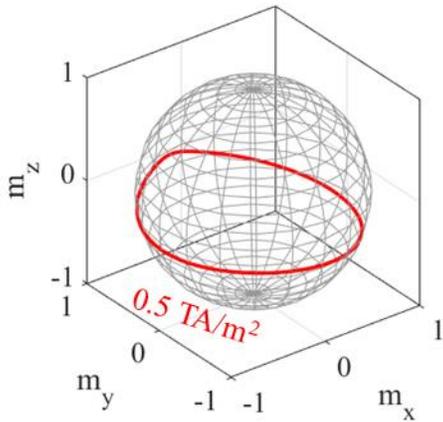

(d)
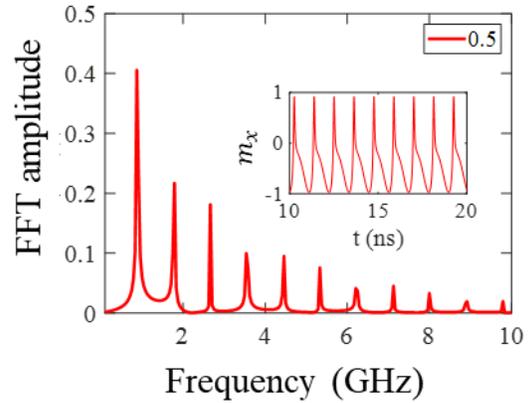

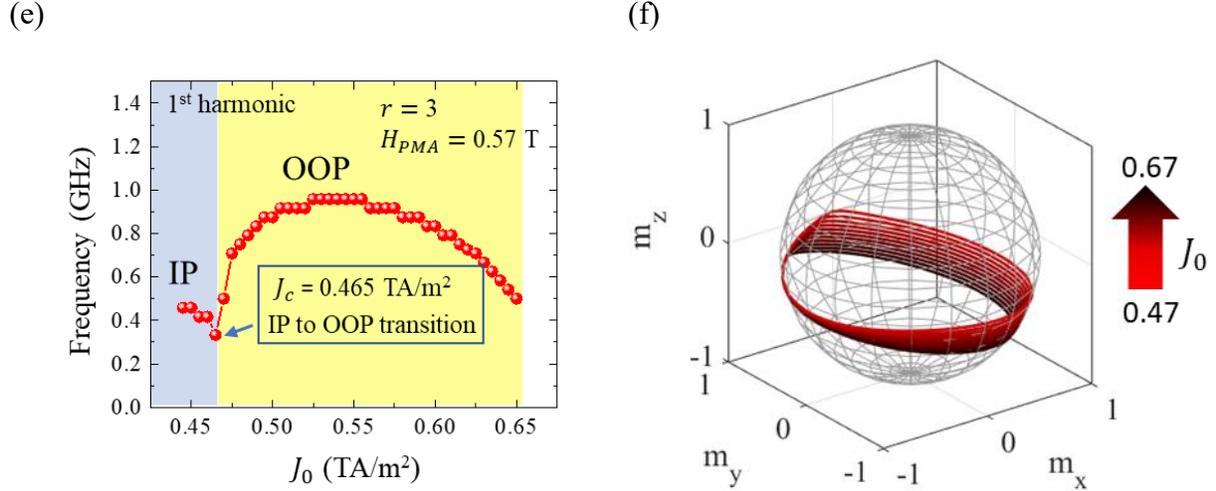

**Figure 2**. (a) Trajectory of the steady state magnetization precession for $J_0 = 0.45$ TA/m² and (b) the corresponding FFT of $m_x(t)$. Similar plots for a higher current density, $J_0 = 0.5$ TA/m² are shown in (c) and (d). The magnetization precession in (a) and (c) represent the IP precession mode and OOP precession mode respectively. (e) Variation of the 1st harmonic of the auto-oscillation frequency as a function of input current density $J_0$, showing the IP to OOP transition in precession mode. (f) Steady state trajectories of OOP precession mode for input current density ranging from 0.45 Ta/m² to 0.67 TA/m².

We further observe in figure 2(e), that the spiking frequency decreases at higher value of $J_0$, as the OOP auto-oscillation trajectory moves away from the $m_z = 0$ plane (see figure 2(f)). However, this leads to generation of better-quality spikes which will be discussed in a later section.

It is important to note that the ellipticity of the FM layer is crucial for the bias-free auto-oscillation of $\boldsymbol{m}$. The elliptic geometry of the FM layer creates a stronger demagnetization field along the minor axis compared to the major axis due to shape anisotropy. In stable equilibrium, the magnetization prefers alignment along the major axis (i.e., along $\boldsymbol{e_x}$), as the PMA is not strong enough to fully compensate for the demagnetization field along the film normal. However, when spin current is injected into the FM layer, the SOT attempts to align the magnetization along the spin polarization direction $\boldsymbol{e_y}$, that coincides with the minor axis. This sets up a competition between the SOT and the in-plane demagnetization field, that eventually triggers the auto-oscillation of magnetization. To examine the impact of this shape anisotropy on auto-oscillation, we varied the axis ratio of the ellipse, while keeping the area constant. This approach ensures a constant total spin current injection into the FM for a particular value of $J_0$, regardless of the axis ratio. Figure 3 shows the 1st harmonic frequency as a function of input current density for different values of axis ratio, $r$. It reveals two key observations: first, the threshold current density increases with increasing axis ratio, and second, the OOP precession is preferred at higher axis ratios. These observations highlight the significant influence of shape anisotropy on the origin and characteristics of auto-oscillation.

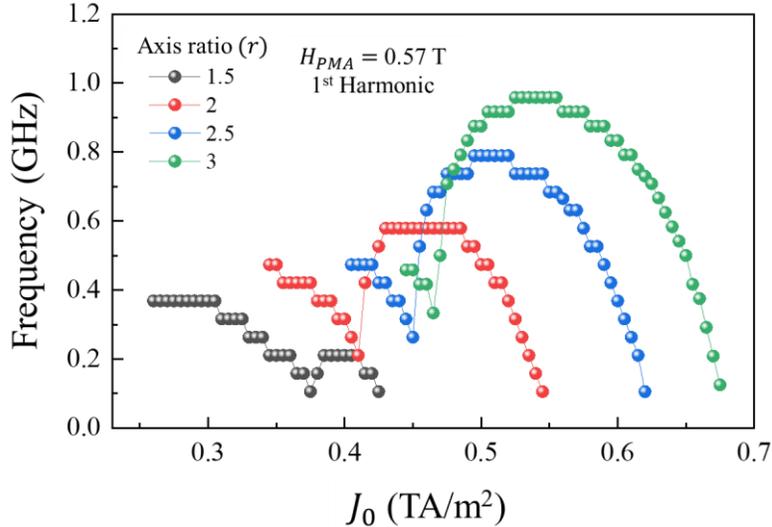

**Figure 3**. Variation of the 1st harmonics of auto-oscillation frequencies as a function of input current density ($J_0$), for different values of axis ratio ($r$).

The above results encourage us to gain a deeper insight into the origin of bias field-free auto-oscillation in our SHNO. However, the microscale physics becomes more complex to understand in the micromagnetic regime. In contrast, the macrospin theory can offer a much simpler model to understand the essential dynamics of the system. In the following section, we investigate the origin of the bias-field free auto-oscillation in our SHNO using the macrospin theory.

## IV. ORIGIN OF THE BIAS FIELD-FREE AUTO-OSCILLATION: A MACROSPIN PERSPECTIVE

To understand the origin of bias-free auto-oscillation, we proceed to theoretically calculate the threshold current for exciting the IP auto-oscillation in our system. We employ the macrospin theory along with a linearized LLG equation following the approach of T. Taniguchi[41,43]. In absence of SOT, the orientation of magnetization at stable equilibrium is determined by the minimum energy state. The energy density is given as $E = -M_s \int d\mathbf{m}.\mathbf{H'}_{eff}$, where $\mathbf{H'}_{eff} = \mathbf{H}_{Oe} + \mathbf{H}_{dem} + \mathbf{H}_{PMA}$, denotes the effective magnetic field experienced by the magnetization. Hence, the energy density for our system can be written (excluding the constant term) as

$$E = -M_s H_{Oe} \sin\theta \sin\phi - \frac{M_s H_A}{2}\sin^2\theta \cos^2\phi + \frac{M_s H_d}{2}\cos^2\theta \quad (3)$$

Here, $H_{Oe} = -\mu_0 J_0 t_{HM}/2$, $H_A = \mu_0 M_s(N_y - N_x)$ and $H_d = \mu_0 M_s(N_z - N_y) - H_{PMA}$.

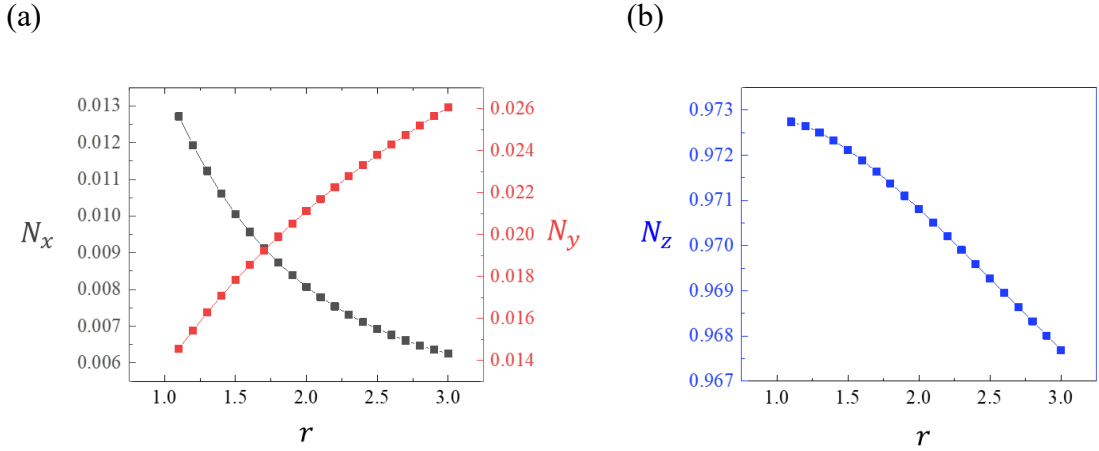

**Figure 4**. Variation of the demagnetization coefficients, $N_x$, $N_y$ (a) and $N_z$ (b) along x, y and z-axes respectively as a function of axis ratio, $r$.

Note that, the demagnetization coefficients $N_x, N_y$ and $N_z$, are strongly dependent on the axis ratio, $r$ (see figure 4), as calculated analytically[34]. $\theta$ and $\phi$ in equation 2 denote the zenith and azimuthal angle of $\mathbf{m}$ such that $\mathbf{m} = (\sin\theta\cos\phi, \sin\theta\sin\phi, \cos\theta)$. The minimum energy state then corresponds to $\theta = \theta_0 = 90°$ and $\phi = \phi_0 = \sin^{-1}(H_{Oe}/H_A)$, which locates close to the major axis of the ellipse in xy-plane (see supplementary information for derivation). In presence of spin current, the SOT destabilizes the magnetization from the minimum energy state. If the SOT is strong enough to compensate for the damping, it can eventually excite IP auto-oscillation. Assuming a small oscillation amplitude at the onset of magnetization auto-oscillation around the equilibrium stationary point, we can expand the LLG equation around the minimum energy state keeping only the first order terms in the dynamic $\boldsymbol{m}(t)$. From such linearized LLG equation, the instability condition can be extracted as[41]

$$H_{SOT}(\boldsymbol{e_y}.\widehat{\boldsymbol{m_0}}) = \frac{\alpha(H_X + H_Y)}{2} \tag{4}$$

Here, $H_{SOT} = \frac{\hbar \theta_{SH} J_0}{2eM_s t_{FM}}$, $H_X = H_{Oe} \sin \phi_0 + H_A \cos^2 \phi_0 + H_d$ and $H_Y = H_{Oe} \sin \phi_0 + H_A \cos 2\phi_0$. Now we rewrite $H_{Oe} = -k_1 J_0$ and $H_{SOT} = k_2 J_0$ for convenience of notation where, $k_1 = \frac{\mu_0 t_{HM}}{2}$ and $k_2 = \frac{\hbar \theta_{SH}}{2eM_s t_{FM}}$. Hence, from equation 3, we finally obtain (see supplementary for the derivation):

$$J_{th} = \sqrt{\frac{\alpha H_A (2H_A + H_d)}{\alpha k_1^2 - 2k_1 k_2}} \tag{5}$$

From the above equation, we find that $J_{th}$ will have a real value only if, (i) $k_1 \neq 0$ i.e., $H_{Oe} \neq 0$ and (ii) $k_2 < 0$ which implies $\theta_{SH} < 0$. Hence, we conclude that, firstly, the Oersted field is necessary to achieve the field-free oscillation in our system, and secondly, we must choose the HM material with a negative spin Hall angle such as $\beta$-W. In fact, without the Oersted field, the equilibrium magnetization points along the major axis which is orthogonal to the spin polarization direction. Therefore, in that case, the SOT cannot destabilize the magnetization. Hence, the $H_{Oe}$ acts as the symmetry breaking field in our system.

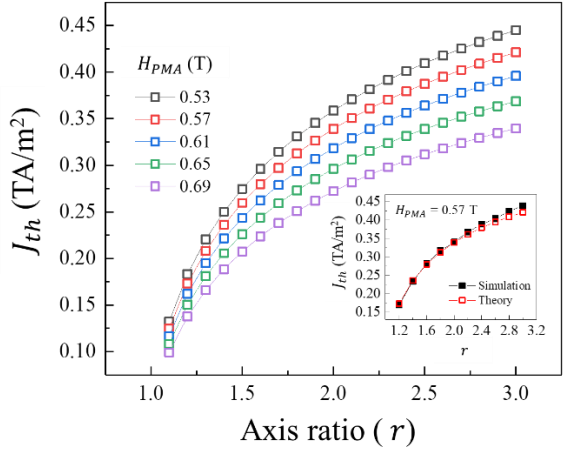

**Figure 5**. Varitaion of the threshold current density ($J_{th}$) as a function of axis ratio ($r$), calculated using macrospin theory for different values of PMA field ($H_{PMA}$). (Inset) Comparison of $J_{th}$ calculated analytically using macrospin model and from micromagnetic simulation, for $H_{PMA} = 0.57$ T.

Figure 5 shows the behavior of $J_{th}$ as a function of $r$ for different values of $H_{PMA}$. Higher value of the PMA field leads to lower $H_d$, causing a reduction in $J_{th}$ as evident from both the equation 4 and the main plots in figure 5. Additionally, the inset of figure 5 demonstrates good agreement between the simulated values of $J_{th}$ and those obtained from equation 4, confirming the validity of the linearized LLG equation for determining the instability threshold in our system.

## V.  SPIKING BEHAVIOR OF SHNO

### A. IP to OOP transition in the auto-oscillation trajectory

It was observed (figure 2a) that, the magnetization undergoes bias field-free auto-oscillation in IP precession mode if the current density is less than a critical value, denoted by $J_c$. Above the critical current density, the magnetization precession becomes OOP (figure 2c). Additionally, comparing figure 2(e) and figure 3, one can see that, $J_c$ is directly dependent on the ellipticity (axis ratio, $r$) of the FM layer. Hence, one might be interested to obtain $J_c$ as a function of $r$. Note that, the linearized LLG equation will no longer be valid because of the large precession amplitude of $\boldsymbol{m}$. In fact, obtaining a simple analytical expression for $J_c$ as a function of $r$, is practically impossible for our system without the biasing field. The difficulties are well discussed in ref. [41] and in the supplementary information. However, a qualitative picture can be drawn as follows: at lower value of $r$, out-of-plane demagnetization field dominates which favors IP precession. In contrast, higher value of $r$ leads to more geometrical confinement along y-axis that results in higher demagnetization field along y-axis and decrease in out-of-plane demagnetization field. Therefore, OOP precession is dominant for higher axis ratio. Hence, from figure 3 we observe that, the oscillation is mostly IP for $r = 1.5$, whereas the magnetization precession is predominantly OOP for majority of the input current density range in case of $r = 3$. Additionally, considering only the shape anisotropy and PMA, macrospin model shows that $\Delta J_c/\Delta r > 0$ (see supplementary information). This is consistent with our micromagnetic simulation results, where $J_c$ increases with $r$ (see figure 3).

## B. Origin of spiking in OOP precession

We now focus on understanding the spiking behavior in our bias-free SHNO in OOP precession mode. In SHNO, the sustained precession of magnetization occurs because of the interplay between the energy provided by SOT and the energy dissipation through damping. The work done by SOT ($W_{SOT}$) and the damping torque ($W_\alpha$) in a full oscillation cycle, are given as[43]:

$$W_{SOT} = \gamma M_s H_{SOT} \oint [\boldsymbol{p}.\boldsymbol{H}_{eff} - (\boldsymbol{m}.\boldsymbol{p})(\boldsymbol{m}.\boldsymbol{H}_{eff})]\, dt, \qquad (6)$$

$$W_\alpha = -\gamma \alpha M_s \oint [H_{eff}^2 - (\boldsymbol{m}.\boldsymbol{H}_{eff})^2]dt \qquad (7)$$

As the input current density exceeds the threshold value, the energy supplied to the magnetization by SOT surpasses the energy dissipation through damping. However, because of the large amplitude precession in our SHNO, the SOT opposes the damping only at certain points on the trajectory of magnetization precession and pumps in power to the oscillatory magnetization. In contrast, the SOT enhances the damping at other points on the precession trajectory resulting in power loss in the system. This behavior of SOT is directly reflected in the time evolution of the power transferred by SOT $\left(\frac{dW_{SOT}}{dt}\right)$ and the power dissipation by intrinsic damping torque $\left(\frac{dW_\alpha}{dt}\right)$, as shown in figure 6(a). Hence, we observe in figure 6(a) that $\frac{dW_{SOT}}{dt}$ (shown in red) is positive for part of the oscillation cycle and negative for the rest of the cycle. Conversely, $\frac{dW_\alpha}{dt}$ (shown in blue) is always negative. We further observe from figure 6(a) that the dynamic fluctuation of the power associated with SOT occurs in spiking manner. This can be qualitatively explained considering the interplay of SOT, Oersted field and the shape anisotropy due to elliptic geometry of the FM layer. In the auto-oscillation regime of magnetization dynamics, as the SOT orients the magnetization along the spin polarization direction ($e_y$), the shape anisotropy (along x-axis) and the Oersted field (along $-e_y$) pull the magnetization away from the spin polarization direction. The SOT, therefore, keep on adding more power to the dynamic magnetization to orient it along $e_y$ again. This leads to accelerate the magnetization towards $e_y$ along the precession trajectory. As a result, the SOT draws off power from the dynamic magnetization by enhancing the effective damping and reduce the precessional angular momentum to orient the magnetization along $e_y$. This cycle repeats again which leads to the spiking nature of $\frac{dW_{SOT}}{dt}$.

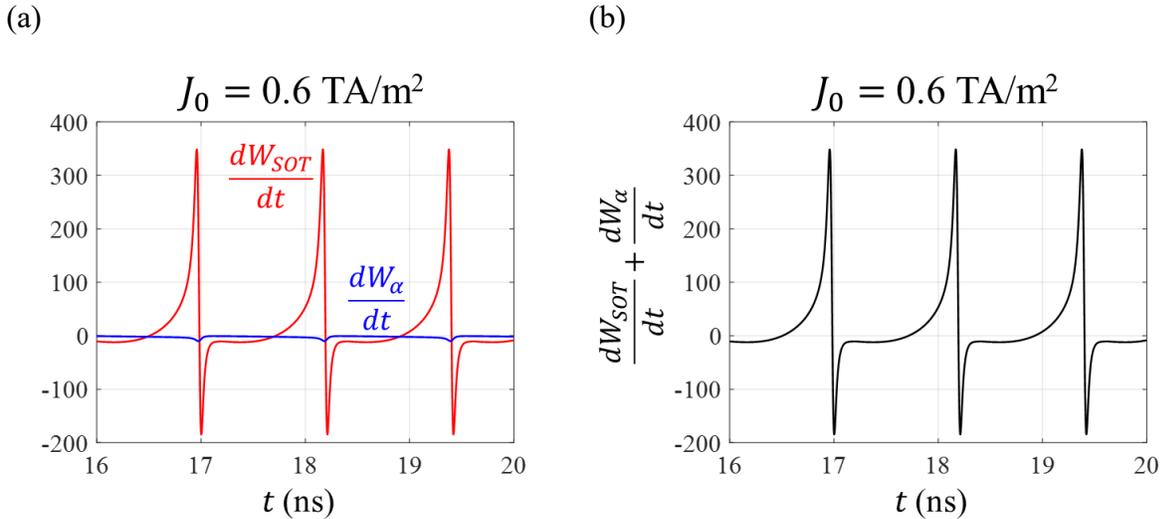

**Figure 6**. Dynamics of the power associated with SOT and damping torque for input current density, $J_0 = 0.6$ TA/m². (a) Time evolution of the power associated with SOT (red) and intrinsic damping torque (blue). (b) Net power available to the system as a function of time (the time independent part is ignored).

Figure 6(b) shows the time evolution of the net power available to the system $\left(\frac{dW_{SOT}}{dt} + \frac{dW_\alpha}{dt}\right)$ for $J_0 = 0.6$ TA/m², which follows the time evolution of $\frac{dW_{SOT}}{dt}$ and fluctuates between negative (net power loss) and positive (net power gain) values throughout the oscillation period. The SOT-driven magnetization precession is, therefore, majorly dictated by the additive or negative power transfer through SOT in a spiking manner. This causes the periodic spiking behavior in the self-oscillatory magnetization dynamics in our SHNO, driven by a constant dc current.

## C. Tunability of spiking through input current

As mentioned earlier, the spiking occurs in the OOP precession state, with the spiking rate given by the 1$^{st}$ harmonic extracted from the FFT of $m_x$. The current-tunability of the spiking rate has been shown in figure 7(a). In course of precession, the SOT tries to align $m$ along the spin-polarization direction $e_y$, which is opposed by the demagnetization field along y-axis. Hence, a stronger demagnetization field along y-axis make $m$ move faster to pass the $e_y$ direction. Conversely, a lower value of demagnetization field along y-axis, allows $m$ stay oriented along $e_y$ for a longer duration, resulting in a lower auto-oscillation frequency and spiking rate. Figure 7(b) shows the variation of the time-averaged y-component of demagnetization field, which is opposite to the behavior of spiking rate shown in figure 7(a).

(a) 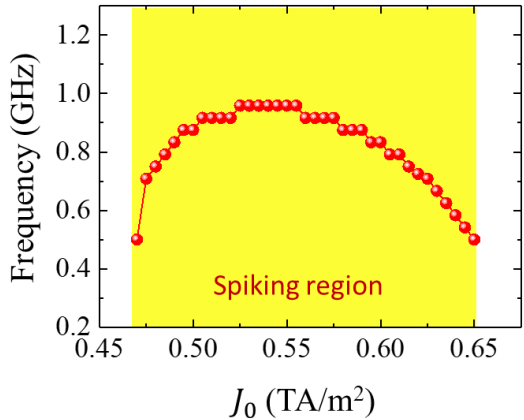  (b) 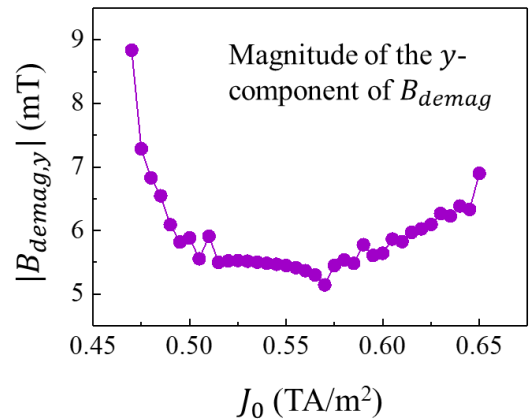

**Figure 7**. (a) Variation of spiking rate (1$^{st}$ harmonic of auto-oscillation frequency) as a function of input current density, $J_0$ in OOP precession mode. This shows the current-tunability of spiking rate. (b) Variation of the magnitude of y-component of demagnetization field, $B_{demag}$ as a function of $J_0$.

Our bias field-free SHNO exhibits not only current tunability of spiking rate but also improved spike sharpness with higher input current density. As shown in figure 2(f), at higher input current, the precession trajectory becomes more out-of-plane. Figure 8 illustrates the variation of the time-averaged value of $\theta$, measured from the z-axis (see the inset of figure 8), as a function of input current density. Spiking in $m_x(t)$ for certain values of time-averaged $\theta$ are also shown in figure 8. Note that as the $\theta$ becomes relatively smaller i.e., $\boldsymbol{m}$ moves closer to the z-axis at higher input current, sharper spikes are generated. This highlights the tunability of spike-quality by varying the input current.

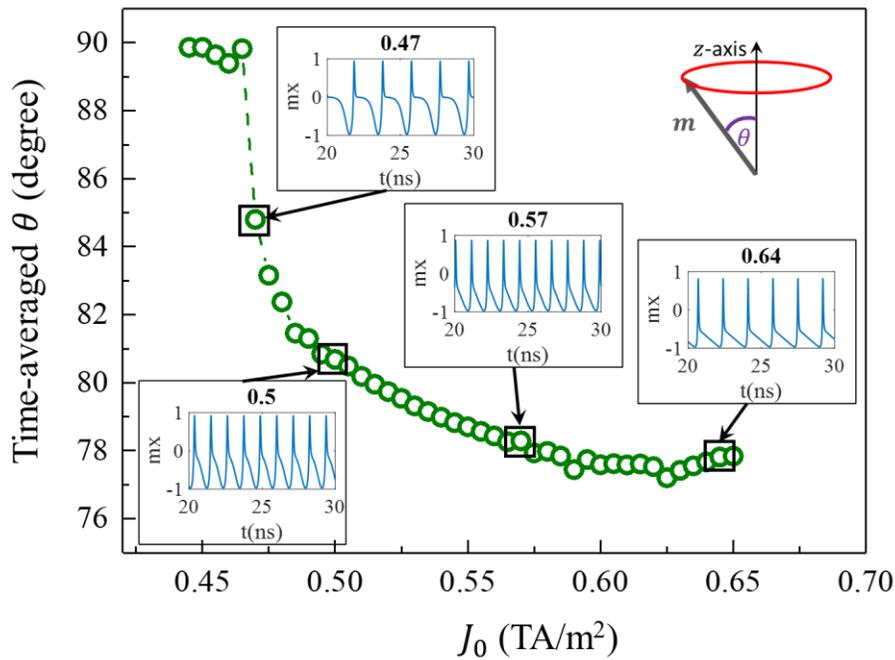

**Figure 8**. Variation of the time-averaged value of the out-of-plane angle, $\theta$ (measured from the z-axis) as a function of input current density, $J_0$. Sharper spikes are obtained at higher current density and lower value of $\theta$. This indicates the current-tunability of spiking rate and quality.

Finally, we demonstrate a leaky-integrate-and-fire (LIF) behavior and reconfigurable spiking behavior in our bias field-free SHNO. In LIF neuron model, the consecutive input excitatory pulses gradually increase the membrane potential while it decreases (leaks) in absence of the excitatory pulses. After a certain number of consecutive input excitatory pulses, the membrane potential surpasses a threshold, and the neuron fires. A similar behavior has been obtained in our SHNO when it is excited by consecutive sub-nanosecond current pulses in presence of a constant biasing current density less than $J_{th}$. Figure 9(a) depicts this LIF behavior of our SHNO. Note that the firing threshold is achieved when $m_x$ value is slightly higher than 0.5. This LIF behavior highlights the potential of such bias field-free SHNO for neuromorphic applications such as image recognition using spiking neural network[44]. In figure 9(b), we show the reconfigurable spiking behavior of our SHNO upon excitation by a linearly varying input current density. In spiking neural network, reconfigurable spiking neurons are designed to dynamically adapt to the variation in input and adjust their properties during runtime. As shown in figure 9(b), our bias-field-free SHNO exhibits on-the-fly variation in spiking rate, spike-amplitude, and spike-sharpness in response to varying input current density. This adaptability would allow the SHNO to efficiently process time-varying and event-based data, enhancing its flexibility and performance with changing inputs. However, this reconfigurability is accomplished mostly at higher input currents as seen in figure 9(b). Both these behaviors strongly suggest the potential of such bias free SHNO in designing efficient neuromorphic hardware with spiking neurons.

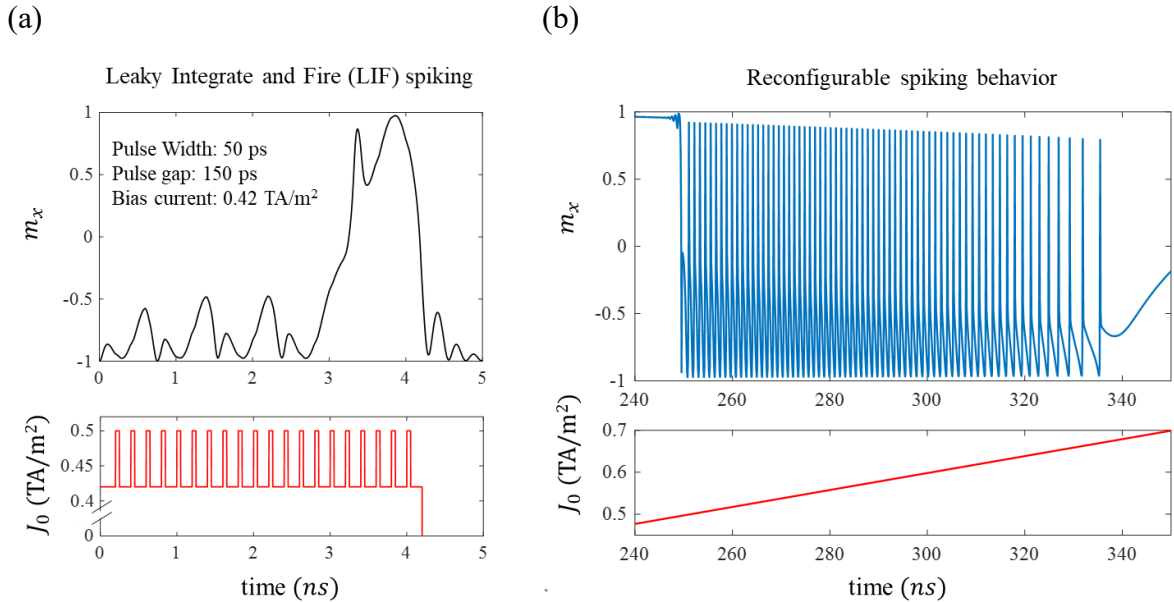

**Figure 9**. (a) Leaky-integrate-and-fire (LIF) behavior in SHNO in presence of consecutive excitatory pulses along with a constant bias current. (b) Reconfigurable spiking behavior in SHNO in presence of a linearly varying input current density. At relatively higher values current density, the spiking rate and quality are self-adjusted based on the value of input current density.

## VI. CONCLUSION

In conclusion, this study unveils intriguing insights into the behavior of bias-free SHNO with simple elliptic geometry and its potential applications in designing artificial neurons for SNN based neuromorphic hardware. In addition to the bias field-free auto-oscillation of magnetization, we have achieved a current-tunable transition from IP to OOP precession mode. In the OOP precession mode, our SHNO exhibits neuron-like spiking dynamics which can be modulated in terms of spiking rate and spike-sharpness through adjustments in input dc current. Notably, we observed both periodic and LIF spiking behavior in the same SHNO device, achievable by manipulating the pulse width and amplitude of the input dc current. Furthermore, the large precession amplitude ensures that the output spike signals are sufficiently strong for efficient signal processing. Our discussion on the impact of geometrical parameters on determination of auto-oscillation characteristics will help to optimize the geometry in a much more efficient way for greater scalability. Therefore, this research can motivate to further explore the potential of bias field-free SHNOs in next generation computing and information processing paradigms.

**SUPPLEMENTAL MATERIAL**

See the Supplementary Information for additional details.

**AUTHOR CONTRIBUTION**

S. M. conceived the original idea. S. M. performed all the micromagnetic simulation. S. M. did all the analysis of the simulation results. R.M. and R. S. R. reviewed the manuscript and provided valuable inputs. R. S. R. supervised the project.

## ACKNOWLEDGEMENT

S.M. would like to acknowledge the NTU-Research Scholarship (NTU-RSS). R.S.R. acknowledges the research project support by the Ministry of Education, Singapore, under its Academic Research Tier 1 grant number RG76/22. Any opinions, findings and conclusions or recommendations expressed in this material are those of the author(s) and do not reflect the views of the Ministry of Education, Singapore.

## DATA AVAILABILITY

The data that support the findings of this study are available from the corresponding author upon reasonable request.

## CONFLICT OF INTEREST

The authors declare no conflict of interest.

## REFERENCES

[1]  J. Grollier, D. Querlioz, K. Y. Camsari, K. Everschor-Sitte, S. Fukami, and M. D. Stiles, Nat Electron **3** (2020).
[2]  D. V. Christensen *et al.*, Neuromorphic Computing and Engineering **2**, 022501 (2022).
[3]  T. Taniguchi, S. Tsunegi, S. Miwa, K. Fujii, H. Kubota, and K. Nakajima, in *Reservoir Computing: Theory, Physical Implementations, and Applications*, edited by K. Nakajima, and I. Fischer (Springer Singapore, Singapore, 2021), pp. 331.
[4]  F. C. Hoppensteadt and E. M. Izhikevich, Physical Review Letters **82**, 2983 (1999).
[5]  T. Aonishi, K. Kurata, and M. Okada, Physical Review Letters **82**, 2800 (1999).
[6]  R. Matsumoto, S. Lequeux, H. Imamura, and J. Grollier, Physical Review Applied **11** (2019).
[7]  J. Torrejon *et al.*, Nature **547**, 428 (2017).
[8]  M. Zahedinejad, A. A. Awad, S. Muralidhar, R. Khymyn, H. Fulara, H. Mazraati, M. Dvornik, and J. Akerman, Nat Nanotechnol **15**, 47 (2020).
[9]  M. Zahedinejad, H. Fulara, R. Khymyn, A. Houshang, M. Dvornik, S. Fukami, S. Kanai, H. Ohno, and J. Akerman, Nat Mater **21**, 81 (2022).
[10] D. I. Albertsson, M. Zahedinejad, A. Houshang, R. Khymyn, J. Åkerman, and A. Rusu, Applied Physics Letters **118** (2021).
[11] A. Houshang *et al.*, Physical Review Applied **17** (2022).
[12] A. J. Mathew, J. R. Mohan, R. Feng, R. Medwal, S. Gupta, R. S. Rawat, and Y. Fukuma, IEEE Transactions on Magnetics **59**, 1 (2023).

[13] J. R. Mohan, A. J. Mathew, K. Nishimura, R. Feng, R. Medwal, S. Gupta, R. S. Rawat, and Y. Fukuma, Sci Rep **13**, 7909 (2023).
[14] J. E. Hirsch, Physical Review Letters **83**, 1834 (1999).
[15] V. E. Demidov, S. Urazhdin, A. Zholud, A. V. Sadovnikov, and S. O. Demokritov, Applied Physics Letters **105** (2014).
[16] A. A. Awad, P. Dürrenfeld, A. Houshang, M. Dvornik, E. Iacocca, R. K. Dumas, and J. Åkerman, Nature Physics **13**, 292 (2016).
[17] A. A. Awad, A. Houshang, M. Zahedinejad, R. Khymyn, and J. Åkerman, Applied Physics Letters **116** (2020).
[18] P. Durrenfeld, A. A. Awad, A. Houshang, R. K. Dumas, and J. Akerman, Nanoscale **9**, 1285 (2017).
[19] H. Fulara, M. Zahedinejad, R. Khymyn, A. A. Awad, S. Muralidhar, M. Dvornik, and J. Åkerman, Science Advances **5**, eaax8467.
[20] H. Fulara, M. Zahedinejad, R. Khymyn, M. Dvornik, S. Fukami, S. Kanai, H. Ohno, and J. Akerman, Nat Commun **11**, 4006 (2020).
[21] A. Kumar, M. Rajabali, V. H. Gonzalez, M. Zahedinejad, A. Houshang, and J. Akerman, Nanoscale **14**, 1432 (2022).
[22] H. Mazraati *et al.*, Physical Review Applied **18** (2022).
[23] B. Jiang, W. Zhang, J. Li, S. Yu, G. Han, S. Xiao, G. Liu, S. Yan, and S. Kang, AIP Advances **10** (2020).
[24] B. Jiang *et al.*, Journal of Magnetism and Magnetic Materials **490** (2019).
[25] S. Manna, R. Medwal, S. Gupta, J. R. Mohan, Y. Fukuma, and R. S. Rawat, Applied Physics Letters **122** (2023).
[26] T. Shirokura and P. N. Hai, Journal of Applied Physics **127** (2020).
[27] R. Khymyn, I. Lisenkov, V. Tiberkevich, B. A. Ivanov, and A. Slavin, Sci Rep **7**, 43705 (2017).
[28] H. Bradley, S. Louis, C. Trevillian, L. Quach, E. Bankowski, A. Slavin, and V. Tyberkevych, AIP Advances **13** (2023).
[29] D. Marković, M. W. Daniels, P. Sethi, A. D. Kent, M. D. Stiles, and J. Grollier, Physical Review B **105** (2022).
[30] A. L. Hodgkin and A. F. Huxley, The Journal of Physiology **117**, 500 (1952).
[31] D. Wang *et al.*, Nat Commun **14**, 1068 (2023).
[32] D. R. Rodrigues, R. Moukhader, Y. Luo, B. Fang, A. Pontlevy, A. Hamadeh, Z. Zeng, M. Carpentieri, and G. Finocchio, Physical Review Applied **19** (2023).
[33] X. Liang, X. Zhang, J. Xia, M. Ezawa, Y. Zhao, G. Zhao, and Y. Zhou, Applied Physics Letters **116** (2020).
[34] M. Beleggia, M. D. Graef, Y. T. Millev, D. A. Goode, and G. Rowlands, Journal of Physics D: Applied Physics **38**, 3333 (2005).
[35] W. Skowroński *et al.*, Physical Review Applied **11** (2019).
[36] R. Bansal, G. Nirala, A. Kumar, S. Chaudhary, and P. K. Muduli, SPIN **08**, 1850018 (2018).
[37] Y.-C. Lau and M. Hayashi, Japanese Journal of Applied Physics **56** (2017).
[38] J. C. Slonczewski, Journal of Magnetism and Magnetic Materials **159**, L1 (1996).
[39] L. Liu, T. Moriyama, D. C. Ralph, and R. A. Buhrman, Physical Review Letters **106**, 036601 (2011).
[40] A. Vansteenkiste, J. Leliaert, M. Dvornik, M. Helsen, F. Garcia-Sanchez, and B. Van Waeyenberge, AIP Advances **4** (2014).
[41] T. Taniguchi, Applied Physics Letters **118** (2021).
[42] J. Leliaert, J. Mulkers, J. De Clercq, A. Coene, M. Dvornik, and B. Van Waeyenberge, AIP Advances **7** (2017).
[43] T. Taniguchi and H. Kubota, Physical Review B **93** (2016).
[44] E. Doutsi, L. Fillatre, M. Antonini, and P. Tsakalides, IEEE Transactions on Image Processing **30**, 4305 (2021).

Reconfigurable neural spiking in bias field-free spin Hall nano oscillator

Sourabh Manna[1], Rohit Medwal[2], Rajdeep Singh Rawat[1#]

[1]Natural Science and Science Education, National Institute of Education, Nanyang Technological University, 637616, Singapore.

[2]Department of Physics, Indian Institute of Technology Kanpur, Uttar Pradesh, 208016, India.

# Correspondence rajdeep.rawat@nie.edu.sg.

# SUPPLEMENTARY NOTES

## I.   Micromagnetic Simulation

In our micromagnetic simulation, the magnetization dynamics has been solved exclusively for the FM layer as the nonmagnetic layers are not modelled in Mumax. However, the effect of the HM layer has been considered in the simulation by using the value of spin Hall angle in formulation of SOT. The FM layer has been discretized into a grid of 256 × 128 × 1 rectangular cells with a uniform cell-size of 2 nm × 2 nm × 1.4 nm. The dimension of the cell-size is well below the exchange length of CoFeB, 7.43 nm. To define the elliptic geometry of the FM layer, we have used the "setgeom()" function along with the "ellipse()" function. In addition, the "edgesmooth" function has been used for smoothing of the geometry edge to avoid the staircase effect. We incorporated an absorbing boundary condition to reduce spin wave reflection from the boundary of the elliptic ferromagnetic (FM) layer. This way we could mimic the experimental conditions where irradiated ions increase damping at the boundary during ion-milling process. In our simulation, we defined an absorbing boundary layer (ABL) with a nominal thickness of 5 nm at the ellipse's boundary. This can be achieved by either gradually increasing the damping constant ($\alpha$) as we approach the boundary from the center of the ellipse or by directly setting a high value for $\alpha$ in the ABL. Both approaches effectively emulate the desired absorbing behavior at the boundary.

To excite the SOT induced magnetization dynamics in the FM layer, we have applied a dc spin-polarized current flowing along z-direction. The spin polarization direction has been determined from the orthogonality of the charge current (along x-direction) and spin current (along z-direction). Before solving the LLG equation, the initial magnetization configuration has been obtained by minimizing the total energy of the system. In addition, the spin current was turned on after 4 ns to avoid any transient effect due to initial relaxation of the system. The simulation was carried out for total 40 ns. The auto-oscillation characteristics such as oscillation frequency modes and corresponding amplitude distribution have been extracted from the fast Fourier transform (FFT) of the dynamic reduced magnetization at steady oscillation state, sampled for the last 25 ns.

## II.   Effect of finite temperature

To observe the behavior of our SHNO at room temperature, we carried out the micromagnetic simulations considering T = 300 K. The Mumax3 subroutine deploys a random magnetic field proportional to the ambient temperature to simulate the effect of finite temperature. Figure S1 shows the comparison between the bias field-free auto-oscillation properties at T = 0 K and T = 300 K. We observe the spiking dynamics at finite temperature as well, although the thermal field induces randomness in the spike amplitude (see figure S1 (a) and (b)). Additionally, it is found that the behavior of the 1$^{st}$ harmonic of auto-oscillation in OOP precession mode is more-or-less similar in both cases, although significant thermal noise is present in IP precession mode (see figure S1

(c) and inset). The higher harmonics, present at T = 0 K, are suppressed by the thermal noise because of their low amplitude, as can be seen in the figure S1 (d) and inset.

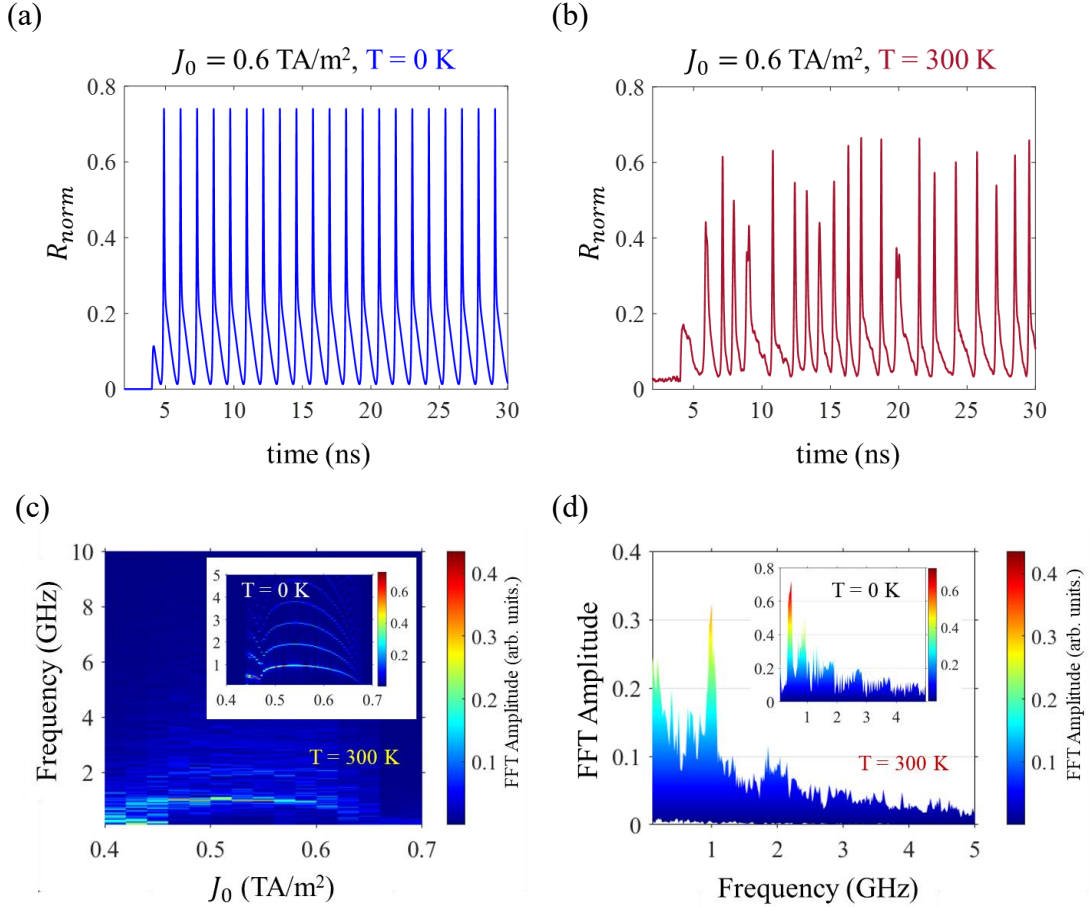

**Figure S1:** Time evolution of normalized TMR, $R_{norm}$ considering 80% TMR ratio simulated at (a) T = 0 K and (b) T = 300 K. (c) Frequency of auto-oscillation as a function of input current density, $J_0$ obtained at T = 300 K and T = 0 K (inset). (d) Variation of FFT amplitude corresponding to the harmonics of auto-oscillation frequency spectra obtained at T = 300 K and T = 0 K (inset).

### III. Derivation of the threshold current density ($J_{th}$) using macrospin theory

We derive the threshold current density of bias field-free auto-oscillation in our SHNO system using macrospin model. The energy density of our elliptic FM, excluding the constant term, can be expressed as (see equation 3 in main text):

$$E = -M_s H_{Oe} \sin\theta \sin\phi - \frac{M_s H_A}{2}\sin^2\theta \cos^2\phi + \frac{M_s H_d}{2}\cos^2\theta \quad (1)$$

Here, $H_{Oe} = -\mu_0 J_0 t_{HM}/2$, $H_A = \mu_0 M_s (N_y - N_x)$ and $H_d = \mu_0 M_s (N_z - N_y) - H_{PMA}$. $\theta$ and $\phi$ in equation 2 denote the zenith and azimuthal angle of **m** such that **m** = $(\sin\theta \cos\phi, \sin\theta \sin\phi, \cos\theta)$. The demagnetization coefficients, $N_x$, $N_y$ and $N_z$ are calculated analytically using the approach of M. Beleggia [1]. In absence of SOT, the orientation of magnetization at stable equilibrium is determined by the minimum of $E$. We define $\theta_0$ and $\phi_0$ as the value of $\theta$ and $\phi$ respectively corresponding to the minimum of $E$. Hence, taking the derivative of $E$ w.r.t. $\theta$ and $\phi$ find that,

$$\frac{\partial E}{\partial \theta} = -M_s H_{Oe} \cos\theta \sin\theta - M_s H_A \sin\theta \cos\theta \cos^2\phi - M_s H_d \cos\theta \sin\theta \tag{2}$$

Now for the minimum energy condition,

$$\frac{\partial E}{\partial \theta} = 0 \Rightarrow \cos\theta_0 = 0 \; i.e., \theta_0 = 90°$$

That makes sense because in our FM, $H_{PMA} < \mu_0 M_s N_z$, so the magnetization prefers to stay in the film-plane denoted by the xy-plane (see figure 1a in main text).

For the minimum of $E$, it should also satisfy the following, given $\theta_0 = 90°$:

$$\frac{\partial E}{\partial \phi} = 0$$

$$\Rightarrow -M_s H_{Oe} \cos\phi_0 + M_s H_A \cos\phi_0 \sin\phi_0 = 0$$

$$\Rightarrow \sin\phi_0 = \frac{H_{Oe}}{H_A}$$

Note that, $\phi_0 = 90°$ cannot be a global minimum of $E$, as the y-axis—which coincides with the minor axis of ellipse—is the in-plane hard axis. However, $\phi_0 = 90°$ represents the saddle point in in the variation of $E(\theta, \phi)$, as shown in figure S2(a). In addition, it turns out that $|H_{Oe}| \ll H_A$ for our system. Moreover, we note that $H_{Oe} < 0$, therefore, $\phi_0 < 0$. Hence, the minimum energy density corresponds to the orientation of $m$ close to the x-axis (major axis of ellipse) with a small component along $-y$ direction due to the Oersted field.

To determine the threshold current density for auto-oscillation, we have followed the approach of T. Taniguchi [2,3]. We use a linearized LLG equation considering small oscillation. Expanding the LLG equation around the minimum energy point, the instability condition for our system reads:

$$H_{SOT}(e_y \cdot \widehat{m_0}) = \frac{\alpha(H_X + H_Y)}{2} \quad (3)$$

Here, $H_{SOT} = \frac{\hbar \theta_{SH} J_0}{2eM_s t_{FM}}$, $H_X = H_{Oe} \sin\phi_0 + H_A \cos^2\phi_0 + H_d$ and $H_Y = H_{Oe}\sin\phi_0 + H_A \cos 2\phi_0$. We further define two constants $k_1$ and $k_2$ such that, $H_{Oe} = -k_1 J_0$ and $H_{SOT} = k_2 J_0$ for easy notations. Therefore, $k_1 = \frac{\mu_0 t_{HM}}{2}$ and $k_2 = \frac{\hbar \theta_{SH}}{2eM_s t_{FM}}$. Note that, at stable equilibrium point $e_y \cdot \widehat{m_0} = e_y \cdot (\sin\phi_0\, e_y + \cos\phi_0\, e_x) = \sin\phi_0$. Hence, from equation 3, we find that:

$$\frac{\hbar \theta_{SH} J_0}{\alpha e M_s t_{FM}} \sin\phi_0 = H_x + H_y \quad (4)$$

Now we write $H_x$, $H_y$ and $\sin\phi_0$ in equation 4 in terms of $H_{Oe}$, $H_A$ and $H_d$. After a few simple algebraic steps, equation 4 finally reads:

$$H_{Oe}^2 + \frac{2}{\alpha} H_{SOT} H_{Oe} = H_A(2H_A + H_d) \quad (5)$$

Substituting $H_{Oe} = -k_1 J_{th}$ and $H_{SOT} = k_2 J_{th}$ for threshold current density $J_{th}$, equation 5, becomes:

$$k_1^2 J_{th}^2 - \frac{2}{\alpha} k_1 k_2 J_{th}^2 = H_A(2H_A + H_d)$$

$$\Rightarrow J_{th} = \sqrt{\frac{\alpha H_A(2H_A + H_d)}{\alpha k_1^2 - 2k_1 k_2}} \quad (6)$$

### IV. Note on analytical calculation of the critical current density using macrospin theory

It would be interesting to calculate $J_c$, the critical current density denoting the transition from in-plane to out-of-plane precession, similar to the theoretical calculation of $J_{th}$. Due to the significant oscillation amplitude at the relevant current density values (i.e., $J_0 \to J_c$), the linearized LLG equation is invalid. As per the auto-oscillation theory in a ferromagnetic thin film, based on a simple macrospin model, the magnetization traces the path of constant energy density in steady state precession. The out-of-plane precession occurs when the energy density, E, crosses the saddle point energy density, $E_{saddle}$. Hence, in principle the $J_c$ can be calculated from:

$$J_c = \lim_{E \to E_{saddle}} J_0(E) \qquad (7)$$

Note that, $J_0$ is a function of energy density, $E$ through the Oersted field, as evident from equation 1. However, it turns out that, obtaining a simple analytical expression for $J_c$ is impossible for bias field-free auto-oscillation of magnetization in our SHNO, considering the Oersted field. T. Taniguchi has thoroughly discussed these challenges for a similar SHNO geometry. Nevertheless, we can roughly determine the qualitative impact of ellipticity on the critical current, $J_c$ by neglecting the Oersted field. The corresponding profile of $E(\theta, \phi)$ is presented in the figure S2 (a) where, the saddle point has been shown. In that case $J_c$ is given by [3]

$$J_c = \frac{\mu_0 \alpha e M_s t_{FM}}{\pi^2 \hbar \theta_{SH}} \sqrt{H_d(r)[H_A(r) + H_d(r)]} \qquad (8)$$

We calculate the variation of $\Delta J_c/\Delta r$ as a function of $r$, depicted in figure S2 (b). The positive values of $\Delta J_c/\Delta r$ denote that $J_c$ should increase as the axis ratio $r$ increases. A similar behavior is also observed in simulation, as can be seen in figure 3 in the main text.

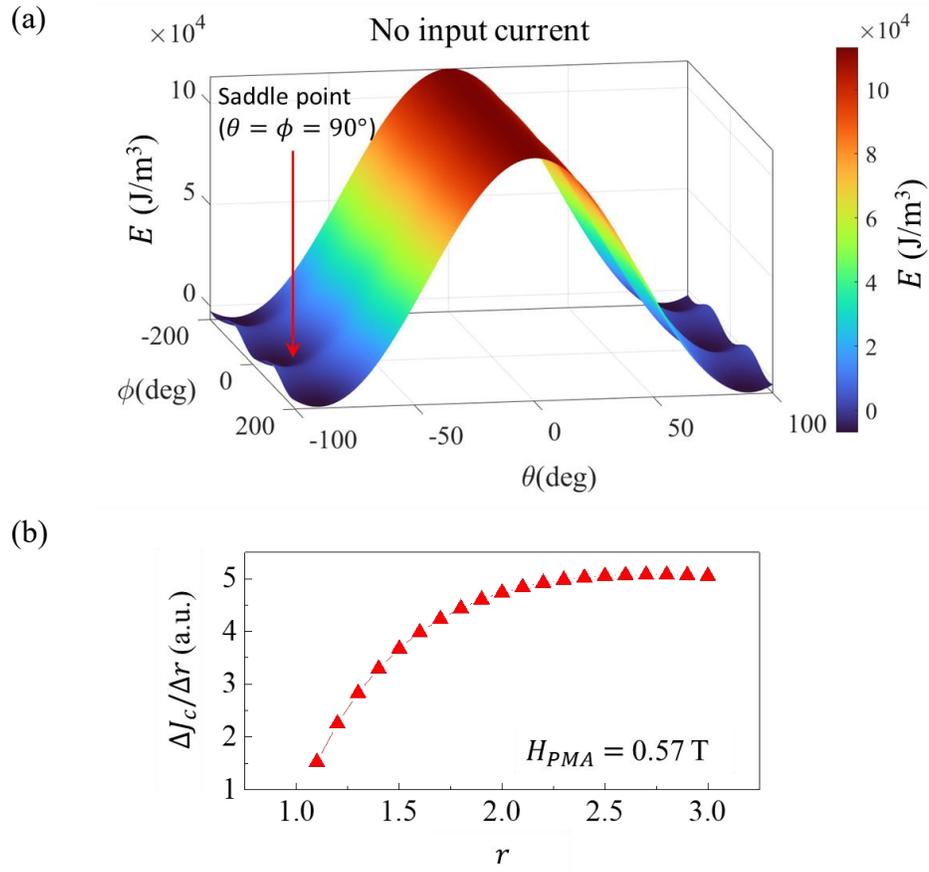

**Figure S2:** (a) Energy density profile as a function of $\theta$ and $\phi$ in absence of input dc current. The saddle energy point obtained for the orientation of $m$ along the minor axis i.e., y-axis, is shown. (b) Variation of $\Delta J/\Delta r$ as function of axis ratio, $r$, calculated from equation 8 in supplementary information.

**References**

1. M. Beleggia, M. D. Graef, Y. T. Millev, D. A. Goode, and G. Rowlands, Journal of Physics D: Applied Physics 38, 3333 (2005).
2. T. Taniguchi and H. Kubota, Physical Review B 93 (2016).
3. T. Taniguchi, Applied Physics Letters 118 (2021).